\documentclass[11pt]{article}
\usepackage{graphicx}
\usepackage{epsfig}
\usepackage{axodraw}
\usepackage{color}

\usepackage{amssymb,amsmath}
\usepackage{graphicx}  
\newcommand{\beq}{\begin{equation}}
\newcommand{\eeq}{\end{equation}}
\newcommand{\be}{\begin{equation}}
\newcommand{\ee}{\end{equation}}
\newcommand{\bea}{\begin{eqnarray}}
\newcommand{\eea}{\end{eqnarray}}

\fontencoding{T1}
\fontfamily{garamond}
\fontseries{m}
\fontshape{it}
\fontsize{13}{15}
\selectfont

\begin{document}

\title{Structure of Light Scalar Mesons from $D_s$ and $D^0$ Non-Leptonic Decays }
\author{L.~Maiani$^a$, A.D. Polosa$^b$, V.~Riquer$^b$, \\
$^a$Dip. Fisica, Universit\`a  di Roma ``La Sapienza'' and INFN, Roma, Italy\\
$^b$INFN, Sezione di Roma, Roma, Italy}
\date{\today}
\maketitle
\begin{abstract} 
Non-leptonic $D$ meson decays  may provide a reliable testbed for the 
multiquark interpretation of light scalar mesons. 
In this letter we consider $D_s$ decay and show that a 4-quark  $f_0(980)$ meson could induce a decay pattern, which is  forbidden for a $q\bar q$ constituent 
structure. Experimental tests to probe such possibilities are within
reach in the near future.
\newline\newline
\newline
{\bf Preprint No.} Roma1-1449/2007 \newline
\end{abstract}
\section{Introduction}
Recent observations of the three body non-leptonic decays of charmed mesons and of the J$/\psi$ strong decays in similar channels are giving important information on low-energy meson dynamics. We refer in particular to the $\pi\pi$ and $K\pi$ mass distributions in D decays by E791~\cite{E791}, BaBar~\cite{BaBar1}, and to the similar distributions in J$/\psi \to \phi \pi\pi$ and $\phi K\bar K$ by BES~\cite{BESII}. 

In this paper, we analyse the S-wave amplitudes of D$_s$ decays:
\bea
\label{Kcharg}
&&D_s \to \pi^+ \rm{(}K^+ K^-\rm{)}\\
\label{Kneutr}
&&D_s \to\pi^+ \rm{(}K_S K_S\rm{)}\\
\label{pipi}
&&D_s \to\pi^+ \pi^+\pi^-
\eea
which can be observed by BaBar and Belle, following the study of the three Kaon decays of D$^0$ already published in~\cite{BaBar1}:
\be
D^0\to K_S K^+ K^-
\label{D03K}
\ee
In high statistics experiments, the S-wave amplitudes in ~(\ref{Kcharg}) to (\ref{D03K}) can be obtained from the Dalitz plot distributions~\cite{BaBar1} and are expected to be dominated by scalar meson exchange, $f_0$ and $a_0$ in particular.
  
We point out that the observation of the $f_0(980)$ structure in the S-wave amplitudes of (\ref{Kcharg}) and (\ref{Kneutr}) offers a unique possibility to elucidate the valence quark composition of the light scalar mesons. The ratio of decay rates (\ref{Kneutr}) to (\ref{Kcharg}) is  predicted to be close to 1/2 if  $f_0(980)$ is a $I=0$, $q\bar q$ state, while a tetraquark composition, $f_0=[sq][\bar s\bar q]$ $(q=u, d)$~\cite{4qmesons}, could give a different result due to possible interference between $I=0$ and $I=1$ amplitudes. We show that, by fixing one single constant, the ratio of the two isospin couplings, the neutral $K_SK_S$ rate could be made almost to vanish in the accessible region. 

The tetraquark nonet of scalar mesons is supposed to be completed by the $f_0(600)$(=$\sigma$)~\cite{E791, sigma} and by the $\kappa(800)$~\cite{kappa} resonances and we comment as well on the decays:
\bea
&&D_s\to \pi f_0(600) \to 3\pi
\label{Dsigma} \\
&&D_s\to \kappa \bar K, \bar \kappa K\to \pi K\bar K
\label{Dkappa}
\eea

Concerning decays (\ref{D03K}), we explain the remarkable equality of the $K\bar K$ mass spectra for the charged and neutral combinations, $K_SK^+$ and $K^-K^+ $, as due to the rapid opening of the kaonic decay channels of $f_0$ and $a_0$ above the $K\bar K$ threshold, a mechanism pointed out long ago by Flatt\`e {\it et al.}~\cite{flatte}. 

The results of the present work are summarized as follows.
\begin{itemize}
\item The observation of a ratio much smaller than 1/2 for the S-wave rates (\ref{Kneutr}) to (\ref{Kcharg}) would provide a clear demonstration of the non-conventional nature of $f_0$ and $a_0$ mesons.

\item Once the ratio of the rates in (\ref{Kneutr}) and (\ref{Kcharg}) is given,  one is led to a univocal prediction of the S-wave decay rates of (\ref{pipi}) and of :
\be
D_s\to \pi^0 K^+K_S,~\pi^0 \pi^+\eta
\label{Dspi0}
\ee
which we present in this paper, for the case of maximal interference. 

\item Suppression of the $\sigma$ channel (\ref{Dsigma}) is implied by the quark diagrams of Fig.~\ref{3qpairs} while the $\kappa$ band should in principle appear in the Dalitz plot  (\ref{Dkappa}), once the $K^*$ amplitude is subtracted.

\item We find that the experimental distributions of $D^0$ decays, Eq.~(\ref{D03K}),  are consistent with the presence of both $f_0$ and $a_0^+$ in (\ref{D03K}), which is required by the four quark model and gives a consistent picture of the $K \bar K$, S-wave production in D decays, see Fig.~\ref{fig:Swavedist}.
\end{itemize}

We describe resonant amplitudes with a modified Breit-Wigner, according the the prescription of~\cite{flatte}. We use masses and couplings from~\cite{BESII}, for the $f_0$, and from~\cite{CrystBarr} and~\cite{BaBar1} for the $a_0$. 

For tetraquark states, an isospin violating $f_0$-$a_0$ mixing could be induced by the u-d quark mass difference~\cite{massmix}. Data available thus far, however, do not show substantial mixing, see however~\cite{closemix}, and we restrict to exact isospin for simplicity. Generalization of our formalism to the mixed case is straightforward. High statistics data on decays (\ref{pipi}) and (\ref{Dspi0}) would provide a very sensitive determination of the mixing angle.

 \section{Isospin structure of the weak amplitudes}
We consider the (bare) weak hamiltonian corresponding to charm-changing, parity-conserving, Cabibbo allowed decays (\ref{Kcharg},\ref{Kneutr}):
\be
H_W= \frac{G\cos^2\theta_C}{\sqrt{2}}\left[ ({\bar s}\gamma^\mu c)( {\bar u}\gamma_\mu d)~+~(\gamma_\mu \to \gamma_\mu \gamma_5)\right]
\label{heffect}
\ee

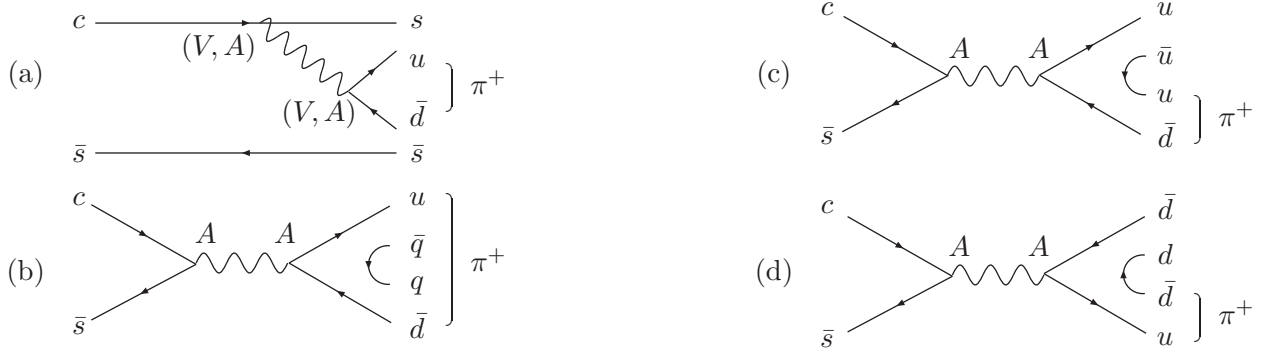
\begin{figure}[ht]
\begin{center}
\fcolorbox{white}{white}{
  \begin{picture}(400,420) (200,-10)
    \SetWidth{1.0}
    
    \SetScale{0.5}
    \SetScaledOffset(100,300) 
    \SetColor{Black}
         \ArrowLine(289,466)(516,466)
    \Photon(413,466)(480,415){7.5}{5}
    \ArrowLine(516,386)(480,414)
    \ArrowLine(480,415)(516,445)
    \ArrowLine(516,368)(289,368)
    \ArrowLine(286,329)(365,284)
    \ArrowLine(365,284)(286,242)
    \Photon(365,285)(434,285){7.5}{3}
    \ArrowLine(512,240)(435,285)
    \ArrowLine(435,285)(511,328)
     \ArrowArc(509.5,283.5)(14.58,84.09,275.91)
    
    \SetScaledOffset(670,592) 
    
    \ArrowLine(361,134)(282,92)
    \ArrowLine(283,179)(362,134)
    \Photon(361,134)(430,134){7.5}{3}
    \ArrowLine(507,90)(430,135)
    \ArrowLine(431,135)(507,178)
    \ArrowLine(286,28)(365,-17)
    \ArrowLine(365,-17)(286,-59)
    \Photon(365,-16)(434,-16){7.5}{3}
    \ArrowLine(434,-15)(511,-60)
    \ArrowLine(510,28)(434,-15)
    \ArrowArc(509.5,134.5)(14.58,84.09,275.91)
    \ArrowArcn(508.5,-15.5)(14.58,275.91,84.09)
     \Text(189,385)[]{{\Black{$ c$}}}
     \Text(317,385)[]{{\Black{$ s$}}}
     \Text(317,370)[]{{\Black{$u$}}}
    \Text(317,350)[]{{\Black{$\bar d$}}} 
    \Text(189,335)[]{{\Black{$\bar s$}}}
     \Text(317,335)[]{{\Black{$\bar s$}}} 
    \Text(189,318)[]{{\Black{$ c$}}}
    \Text(189,270)[]{{\Black{$\bar s$}}}
     \Text(242,375)[]{{\Black{$ (V,A)$}}}
     \Text(280,350)[]{{\Black{$ (V,A)$}}} 

    \Text(237,305)[]{\Black{$ A$}}
    \Text(267,305)[]{\Black{$ A$}}
    \Text(317,318)[]{{\Black{$ u$}}}
    \Text(317,300)[]{{\Black{$\bar q$}}}
    \Text(317,285)[]{{\Black{$q$}}}
    \Text(317,270)[]{{\Black{$\bar d$}}} 
   \Text(472,390)[]{{\Black{$ c$}}}
    \Text(472,342)[]{{\Black{$\bar s$}}}
    \Text(522,375)[]{\Black{$ A$}}
    \Text(552,375)[]{\Black{$ A$}}
    \Text(600,390)[]{{\Black{$ u$}}}
    \Text(600,372)[]{{\Black{$\bar u$}}}
    \Text(600,357)[]{{\Black{$u$}}}
    \Text(600,342)[]{{\Black{$\bar d$}}}
    \Text(472,315)[]{{\Black{$ c$}}}
    \Text(472,265)[]{{\Black{$\bar s$}}}
    \Text(522,300)[]{\Black{$ A$}}
    \Text(552,300)[]{\Black{$ A$}}
    \Text(600,315)[]{{\Black{$ \bar d$}}}
    \Text(600,297)[]{{\Black{$d$}}}
    \Text(600,282)[]{{\Black{$\bar d$}}}
    \Text(600,265)[]{{\Black{$u$}}} 
      \put(329,360){\oval(2,18)[r]}
     \put(329,295){\oval(2,50)[r]}
      \put(612,348){\oval(2,18)[r]}
      \put(612,273){\oval(2,18)[r]} 
      \Text(344,362)[]{$\pi^+$}
       \Text(344,295)[]{$\pi^+$}
        \Text(627,350)[]{$\pi^+$}
       \Text(627,275)[]{$\pi^+$}
       \Text(169,365)[]{(a)}
      \Text(169,290)[]{(b)}
      \Text(452,365)[]{(c)}
       \Text(452,290)[]{(d)}
  \end{picture}
}\end{center}
\vskip-9truecm
\caption{\footnotesize Quark diagrams leading to $\pi^+$ plus one quark-antiquark pair; (a): quark decay. (b,c,d): quark annihilation. Only Axial $\times$ Axial currents contribute to diagrams (b) to (d), since the initial meson is pseudoscalar and the overall transition is parity conserving.}
 \label{2qpairs}
 \end{figure}

Quark diagrams leading to final states with two quark pairs are reported in Fig.~\ref{2qpairs}, and include c-quark decay, (a), and annihilation, (b) to (d). Diagram (a) leads clearly to a $K\bar K$ state with $I=0$ and so does diagram (b), if we assume that the additional quark pair is produced from an isospin invariant, charge-conjugation symmetric sea. 

Diagrams (c) and (d) differ by the exchange $ u\to {\bar d}$ in the weak current and ${\bar d}\to u$ in the sea quarks. Only axial weak currents contribute, due to the pseudoscalar nature of D$_s$ and to parity conservation. 
Therefore, for a symmetric sea, the amplitude takes a plus sign under these exchanges, leading again to an $I=0$ state for $K\bar K$:
\bea
&&A(D_s\to \pi^+ K^+ K^-)=A_a+A_b+A_c=A_a+A_b+A_{AA}\equiv {\cal A}_{I=0}\nonumber\\
&&A(D_s\to \pi^+ K^0 \bar K^0)= A_a+A_b+A_d=A_a+A_b+A_{AA}\equiv{\cal A}_{I=0}
\label{ispinamplqqbar}
\eea

Thus, taking into account the effect due to the difference of $K^+K^-$ and $K^0\bar K^0$ thresholds, the conventional $q\bar q$ picture of the scalar mesons implies uniquely:
\bea
&&\left[\frac {{\rm Rate(}K_S K_S{\rm )}}{{\rm Rate(}K^+ K^-{\rm )}}\right]_{S-wave}=\frac{1}{2}\times \left[\frac {{\rm Rate(}K^0 {\bar K}^0{\rm )}}{{\rm Rate(}K^+ K^-{\rm )}}\right]_{S-wave}\simeq\frac{1}{2}\times 0.83~~~~~~~~~~{\rm (}f_0,a_0=~q\bar q~{\rm )}
\label{qqbar}
\eea

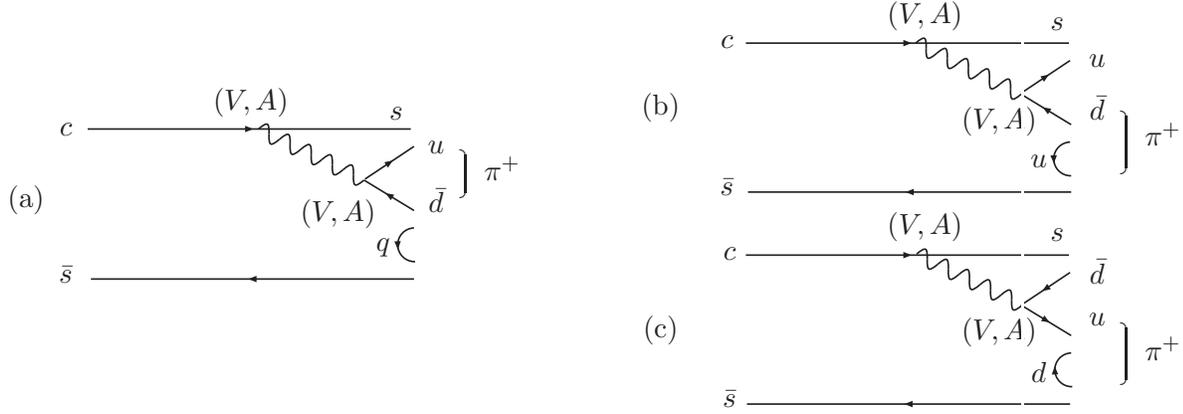
\begin{figure}[ht]
\begin{center}
     \SetScale{0.41}
     
    \SetScaledOffset(250,280)
    
    \SetWidth{1.5}
    \fcolorbox{white}{white}{
     \begin{picture}(273,389) (282,-30)
     \ArrowLine(256,466)(553,466)
    \Photon(413,466)(510,419){7.5}{5}
    \ArrowLine(556,391)(510,419)
    \ArrowLine(511,420)(556,450)
    \ArrowArc(556.5,359.5)(15.51,91.85,271.85)
    \ArrowLine(556,328)(259,328)
    
    \SetScaledOffset(854,555) 
    
    \ArrowLine(257,270)(554,270)
    \Photon(413,270)(510,223){7.5}{5}
    \ArrowLine(510,224)(555,254)
    \ArrowLine(556,195)(510,223)
    \ArrowArc(555.5,163.5)(15.51,91.85,271.85)
    \ArrowLine(556,133)(259,133)
    \ArrowLine(257,75)(554,75)
    \Photon(414,76)(511,29){7.5}{5}
    \ArrowLine(555,59)(510,29)
    \ArrowLine(510,29)(556,1)
    \ArrowArcn(556.5,-30.5)(15.51,271.85,91.85)
    \ArrowLine(555,-62)(258,-62)
    \Text(200,307)[]{{\Black{$ c$}}}
    \Text(200,252)[]{{\Black{$\bar s$}}} 
    \Text(325,313)[]{{\Black{$ s$}}}
    \Text(340,300)[]{{\Black{$ u$}}}
     \Text(340,280)[]{{\Black{$ \bar d$}}}
     \Text(270,317)[]{{\Black{$ (V,A)$}}}
     \Text(303,275)[]{{\Black{$ (V,A)$}}} 
     \Text(320,262)[]{{\Black{$q$}}}      
    \Text(450,340)[]{{\Black{$ c$}}}
    \Text(450,285)[]{{\Black{$\bar s$}}} 
    \Text(575,347)[]{{\Black{$ s$}}}
     \Text(590,333)[]{{\Black{$ u$}}}
     \Text(590,315)[]{{\Black{$ \bar d$}}}  
     \Text(525,350)[]{{\Black{$ (V,A)$}}}  
     \Text(553,310)[]{{\Black{$ (V,A)$}}} 
      \Text(568,295)[]{{\Black{$u$}}}     
    \Text(451,260)[]{{\Black{$ c$}}}
    \Text(451,205)[]{{\Black{$\bar s$}}} 
     \Text(575,267)[]{{\Black{$ s$}}}
     \Text(590,253)[]{{\Black{$ \bar d$}}}
     \Text(590,235)[]{{\Black{$ u$}}}    
     \Text(525,270)[]{{\Black{$ (V,A)$}}}  
     \Text(553,230)[]{{\Black{$ (V,A)$}}}    
     \Text(568,215)[]{{\Black{$d$}}} 
       \put(350,290){\oval(2,18)[r]}
       \put(600,302){\oval(2,23.5)[r]}
       \put(600,222){\oval(2,23.5)[r]}
       \Text(615,305)[]{$\pi^+$}
       \Text(615,224)[]{$\pi^+$}
       \Text(365,292)[]{$\pi^+$}
      \Text(185,280)[]{(a)} %
      \Text(425,315)[]{(b)}
       \Text(425,231)[]{(c)}
  \end{picture}
}
\end{center}
\vskip-8truecm
\caption{\footnotesize Quark-decay diagrams leading to $\pi^+$ plus two quark-antiquark pairs. Contributions from Vector $\times$ Vector and Axial $\times$ Axial currents only, since the overall transition is parity conserving.}
\label{3qpairs}
\end{figure}

The situation is different for the tetraquark structure, where we may expect that quark diagrams with three quark pairs in the final state dominate. The corresponding diagrams are reported in Fig.~\ref{3qpairs}, where we restrict to quark decay amplitudes for simplicity. In this case both axial-axial and vector-vector amplitudes contribute. The latter take a minus sign under the exchange $u\to {\bar d}$. Therefore, we may have both $I=0$ and $I=1$ for the $K\bar K$:
\bea
&&A(D_s\to \pi^+ K^+ K^-)=A_a+A_b=A_a+(A_{AA}+A_{VV})\equiv {\cal A}_{I=0}+{\cal A}_{I=1} \notag \\
&&A(D_s\to \pi^+ K^0 \bar K^0)= A_a+A_c=A_a+(A_{AA}-A_{VV})\equiv{\cal A}_{I=0}-{\cal A}_{I=1}
\label{ispinampl}
\eea
where
\be
{\cal A}_I=\langle (K\bar K)_I; {\rm out}|H_W|D_s\rangle.
\ee

A near cancellation is obtained if the axial-axial and vector-vector amplitudes are close to each other and the amplitude to produce the $\pi^+$ from the weak current is negligible, a not unlikely situation since the latter amplitude vanishes with the pion mass in the free quark limit. 
In the following, we analyze the extreme case where negative interference in the $K_S K_S$ channel is maximal and (see Sect.\ref{analysis}):
\be
\left[\frac {{\rm Rate(}K_S K_S{\rm )}}{{\rm Rate(}K^+ K^-{\rm )}}\right]_{S-wave}\simeq 0~~~~~~{\rm (}f_0, a_0=~[sq][\bar s\bar q]~{\rm )}
\label{4quark}
\ee

\section{Pole model of the S-waves}
\label{pole}

The $K^+K^-$ mass distribution in~(\ref{Kcharg}) is dominated by the $\phi(1020)$. Once the dominant contribution is subtracted, the S-wave amplitude immediately below the $\phi$ region should display the behavior associated with the rising Breit-Wigner of the $f_0(980)$ and $a_0(980)$ toward the peaks, which are however below the $K\bar K$ threshold. The $\phi$ peak is, of course, absent in the $K_S K_S$ mass distribution, which should display the $f_0/a_0$ structure without any subtraction. 

As seen before, the conventional $q\bar q$ picture of the scalar mesons implies the $K\bar K$ system to be in pure $I=0$ state. This needs not to be true if there are two independent isospin amplitudes  for processes (\ref{Kcharg}, \ref{Kneutr}), as it is the case if $f_0$ and $a_0$ are tetraquark states~\cite{4qmesons}.

We dominate the S-wave amplitudes in (\ref{ispinampl}) with the resonant processes:
\bea
&&D_s\to \pi^+ f_0(980); \pi^+ a_0(980);\notag \\
&& f_0/a_0 \to K^+ K^-; K^0 \bar K^0
\eea  

Assuming exact isospin in the mass eigenstates\footnote{we denote by e.g. [su] the fully antisymmetric combination of strange and up quarks~\cite{4qmesons}} we write:
\bea
&&\vert f\rangle= \vert S_0\rangle =\left\vert \frac{[su][\bar u\bar s]+[sd][\bar d\bar s]}{\sqrt{2}} \right\rangle \notag \\
&&\vert a\rangle= \vert S_1\rangle=\left\vert  \frac{[su][\bar u\bar s]-[sd][\bar d\bar s]}{\sqrt{2}}\right\rangle
\label{4qstates}
\eea

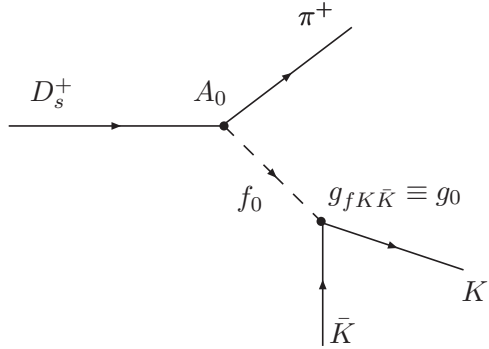
\begin{figure}[ht]
\begin{center}
\vskip2truecm
\SetScale{0.61}
    \SetScaledOffset(220,160)
\fcolorbox{white}{white}{
  \begin{picture}(281,197) (227,-104)
    \SetWidth{1}
    \SetColor{Black}
    \ArrowLine(227,32)(360,32)
    \ArrowLine(360,33)(439,93)
    \DashArrowLine(360,32)(421,-28){10}
    \ArrowLine(421,-104)(421,-28)
    \ArrowLine(421,-28)(508,-57)
    \Vertex(360,32){2.83}
    \Vertex(420,-27){2.83}
    \Text(290,130)[]{{\Black{$ D_s^+$}}}
     \Text(350,130)[]{{\Black{$ A_0$}}}
     \Text(390,160)[]{{\Black{$ \pi^+$}}}
     \Text(390,160)[]{{\Black{$ \pi^+$}}}
      \Text(365,90)[]{{\Black{$ f_0$}}}
      \Text(420,90)[]{{\Black{$ g_{f K \bar{K}}\equiv g_0$}}}
      \Text(450,55)[]{{\Black{$ K$}}}
      \Text(400,40)[]{{\Black{$ \bar{K}$}}}
  \end{picture}
}
\end{center}
\vskip-5truecm
\caption{\footnotesize  Pole diagram for D$^+_s$, S-wave, decay amplitudes. A similar diagram has to be added for $a_0$ exchange. In the notation of Eq.~(\ref{ampls}) of text, $A_0=\langle f\vert A\rangle$ and $g_0 = \langle K\bar K; {\rm out}\vert f\rangle$.}
\label{diagram}
\end{figure}

In an abbreviated notation, we can write the amplitudes of (\ref{Kcharg}) and (\ref{Kneutr}) according to:
\be
A(D_s\to \pi^+K\bar K)=\langle K\bar K;{\rm out} \vert \left(\sum_{I= 0}^1  \vert S_I \rangle BW_I(s) \langle S_I\vert  \right )\vert A \rangle
\label{ampls}
\ee

and we introduce:
\bea
&&g_I=\langle K\bar K;{\rm out}\vert S_I\rangle\\
\label{KKcoupl}
&&A_I=\langle S_I\vert A\rangle
\label{extamp}
\eea



with $I=0,1$. The pole model is described by the diagram in Fig.~\ref{diagram}:
\vskip0.6truecm

BW$_f$(s) and BW$_a$(s) describe the line-shape of the resonances, which we take of the relativistic Breit-Wigner form, e.g.:
\be
BW_f=\frac{1}{s-M_f^2+iM\Gamma_f(s)}
\label{sdepBW}
\ee

M$_f$ and $\Gamma_f$ are the mass and total width. According to Ref.~\cite{flatte}, we introduce an s-dependence in the widths, to take properly into account the opening of the $K\bar K$ threshold inside the $f_0$ or $a_0$ widths. 

The widths that appear in (\ref{sdepBW}) are parameterized as follows.
\begin{itemize}

\item The I =1 charged scalar states, that is the a$_0^\pm$(980), have a distinctive $\eta \pi$ decay:
\be
a_0^\pm(980)\to \eta \pi^\pm
\ee
which, of course, is shared by the I=0 component, $S_1$. Conventionally, the corresponding rate $\Gamma_1$ is written according to:
\be
\Gamma_1=\Gamma(a_0^\pm(980)\to \eta \pi^\pm)=g_{\eta\pi}^2\frac{2p_{\eta\pi}(s)}{\sqrt{s}}
\label{gammaetapi}
\ee
where $p_{\eta\pi}$ is the decay momentum (there would be no real need to introduce an s-dependence here, because the threshold of the final state is quite below the resonance position). In Ref.~\cite{CrystBarr}, $\Gamma_1$ is parameterized with a coupling $(g_1)_{CB}$:
\be
 (g_1)_{CB}^2=M_a g_{\eta\pi}^2
 \label{CBetapi}
 \ee
We obtain from Table~\ref{tab:uno}:
\be
\Gamma_1(M_a)\simeq 69~{\rm MeV}
\label{gaetapi}
\ee

\item The I=0 state, $S_0$, has a distinctive $\pi\pi$ decay. We denote by $\Gamma_0(s)$ the $2\pi$ width and follow the convention of Ref.~\cite{BESII,flatte}:
\be
\Gamma_0(s)=\frac{3}{2}~g_\pi^2  \frac{2~p_\pi(s)}{\sqrt{s}}
\label{ppwidth}
\ee
where the factor 3/2 arises from the charge multiplicity of the final $\pi^+\pi^-$ and $\pi^0\pi^0$ states and $p_\pi$ is the decay momentum. (also in this case the s-dependence is hardly needed). Using Table~\ref{table2}:
\be
\Gamma_0(M_f)\simeq 236~{\rm MeV}
\label{ppwidth2}
\ee
\item $S_{0,1}$ have $K^+ K^-$ and $K^0\bar K^0$ decays characterized by the couplings (\ref{KKcoupl}). We write, e.g. for f$_0$:
\bea
&&\Gamma(f\to K^+K^-)+\Gamma(f\to K^0\bar K^0)=\notag \\
&&=g_0^2\left[  \frac{2p_{\rm ch}(s)}{\sqrt{s}}+ \frac{2p_{\rm neu}(s)}{\sqrt{s}}\right]
\label{f0KK}
\eea
We denote by $p_{\rm ch,neu}$ and $p_{01}$ the decay momenta of $K^+K^-$, $K^0\bar K^0$ and $K^+\bar K^0$ in the scalar particle decays, :
\be
p_{\rm ch,neu}=\frac{1}{2}\sqrt{s-4m_{K^{+,0}}^2};~~~p_{01}=\frac{1}{2\sqrt{s}}\sqrt{\lambda(s,m_{K^+}^2, m_{\bar K^0}^2)}
\label{pK}
\ee
and $\lambda(x,y,z)=x^2+y^2+z^2-2xy-2zx-2zy$, the usual triangular function.

\item We take $g_{0}$ from Ref.~\cite{BESII}:
\be
g_0^2=(g_K^2)_{BES} \simeq 694~{\rm MeV};
\label{g0KKexpt}
\ee
Refs.~\cite{CrystBarr} and~\cite{BaBar1} give different determinations of $g_1$: 
\be
g_1^2=\frac{[(g_2)_{CB}]^2}{2 M_a} \simeq \;\left\{\begin{array} {r@{\quad:\quad}l} 54~{\rm MeV}&{\rm Crystal\;Barrel}\\ 108~{\rm MeV}&{\rm BaBar }\end{array}\right.
\label{g1KKexpt}
\ee
We take the BaBar value for consistency.
\end{itemize}


Summarizing, we give the expressions of the widths for $f_0$, $a_0^0$ and $a_0^\pm$ as follows.
\bea
&&\Gamma_{f_0}=\frac{3}{2}~g_\pi^2  \frac{2~p_\pi(s)}{\sqrt{s}} +g_0^2 \frac{2}{\sqrt{s}}~\left(p_{\rm ch}+p_{\rm neu}\right)\notag \\
&&\Gamma_{a_0^0}=\Gamma_1 +g_1^2~ \frac{2}{\sqrt{s}}\left( p_{\rm ch}+p_{\rm neu}\right)\notag \\
&&\Gamma_{a_0^\pm}=\Gamma_1 +2 g_1^2~ \frac{2~p_{01}(s)}{\sqrt{s}}.
\label{totgam}
\eea
%
\begin{table}[htb]
\caption{{\footnotesize Parameters of the  $a_0^+$, from~\cite{CrystBarr} and~\cite{BaBar1}. The subscript $CB$ refers to the Crystal Barrel definition of the coupling, see Eq.~(\ref{CBetapi}), $(g_{1,2})_{CB}$ refer to the $\eta\pi$ and $K\bar K$ channels, respectively.}}
\begin{center}
\label{tab:uno}
\begin{tabular}{@{}|c|c|c|c|}
\hline
$M_a$~(MeV)~\cite{CrystBarr} & $(g_1)_{CB}$~(MeV)~\cite{CrystBarr}  & $(g_2^2/g_1^2)_{CB}$~\cite{CrystBarr} &  $(g_2)_{CB}$~(MeV)~\cite{BaBar1}\\
\hline
 $999 \pm 2$ & 324$\pm$15 & 1.03$\pm$0.14 & 473$\pm$29$\pm$40 \\
\hline
\end{tabular}\\[2pt]
\end{center}
\end{table}
\begin{table}[htb]
\caption{{\footnotesize Parameters of the  $f_0$, from~\cite{BESII}.}}
\begin{center}
\label{table2}
\begin{tabular}{@{}|c|c|c|}
\hline
$M_f$~(MeV) & $g_\pi^2$~(MeV)  & $g_K^2/g_\pi^2$  \\
\hline
 965$\pm$8 (stat)$\pm$6 (syst) & 165$\pm$10$\pm$15 & 4.21$\pm$0.25$\pm$0.21\\
\hline
\end{tabular}\\[2pt]
\end{center}
\end{table}

The amplitude in (\ref{ampls}), where we use~(\ref{sdepBW}) for the propagator of the unstable particle, corresponds to the Feynman diagram in Fig.~\ref{diagram} (plus the analogous one for $a_0$ exchange). We obtain the following form for the differential distribution in $s=m_{K\bar K}^2$:
\bea
&&\frac{d\Gamma(D_s\to \pi^+K\bar K)}{ds}=\notag \\
&&= \frac{1}{128\pi^3 M_{D_s}^2}
\times \frac{\sqrt{\lambda(M^2_{D_s}, s,m^2_{\pi})}}{2M_{D_s}}\vert A(D_s\to \pi^+K\bar K) \vert^2~\frac{2 p_{K\bar K}}{\sqrt{s}}=|F_{K\bar K}(s)|^2\frac{2 p_{K\bar K}}{\sqrt{s}}
\label{ratepiKK}
\eea
where $p_{K\bar K}$ is given by (\ref{pK}).

With the same normalization of the amplitude, the Dalitz plot density is given by:
\be
\frac{d\Gamma(D\to a b c)} {ds_1 ds_2}=\frac{1}{256\pi^3 M_{D}^3} 
\vert \left. A(D\to abc)\right|_{^{(ab)\to s_1}_{(bc)\to s_2}}\vert^2
\label{Dpdensity}
\ee

\section{D$_s\to(\pi K \bar K)_{S-wave}$ and D$_s\to(\pi^0\pi^+\eta)_{S-wave} $}
\label{analysis}

Above threshold, the $K\bar K$ widths grow rather quickly. One expects~\cite{flatte} the line shape to be dominated by the $K\bar K$ widths and therefore 
a quite similar behavior for $f_0$ and  $a_0$. 

Therefore, it is sufficient to tune the ratio:
 \be
 Q=\frac{A_0}{A_1}
 \label{condition}
 \ee
 to suppress almost completely the $K^0\bar K^0$ amplitude in the accessible region, independently from the fact that ${\cal A}_{I=0}$ and ${\cal A}_{I=1}$ are dominated by different (close lying) resonances. 

We analyze the decays on the basis of Eq.~(\ref{ratepiKK}).
We take the values of the parameters  for $f_0$ from~\cite{BESII} and for $a_0^\pm$ from~\cite{CrystBarr,BaBar1}, as summarized in Tables~\ref{tab:uno} and \ref{table2} .

\begin{figure}[htb]
\begin{minipage}[t]{80mm}
\includegraphics[width=1.0\textwidth]{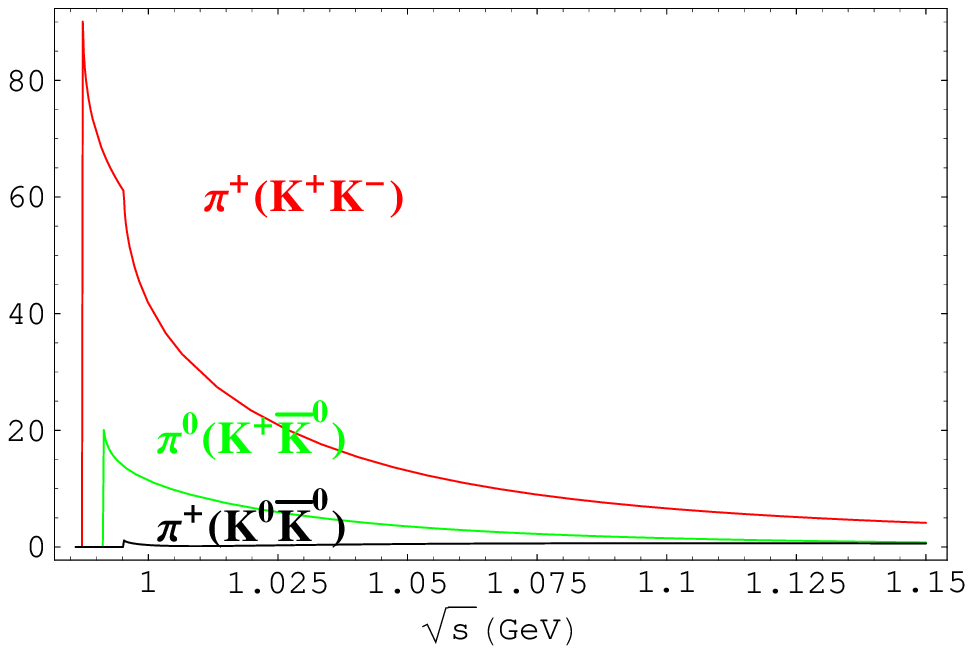}
\end{minipage}
%
%
\begin{minipage}[t]{75mm}
\includegraphics[width=0.98\textwidth]{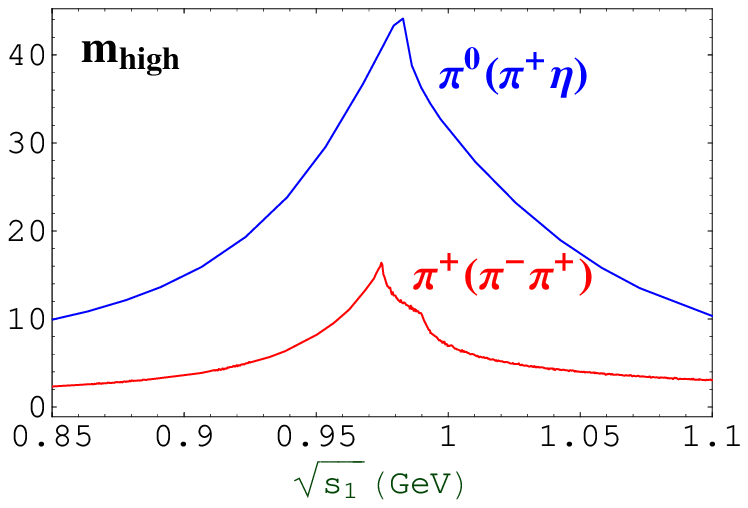}
\end{minipage}
\caption{\footnotesize {\it Left panel}. Phase-space corrected, S-wave probabilities, $|F|^2$ see Eq.~(\ref{ratepiKK}), for the decay modes $D_s^+\to\pi K{\bar K}$ as function of the invariant mass of the systems in parentheses. The value $Q=A_0/A_1=1.5$ is assumed. {\it Right panel}. Distribution of the rates of $D_s \to \pi^+(\pi^+\pi^-)_{h,l}$ (red) and $D_s \to \pi^0(\pi^+\eta)$ (blue). The $\pi^+(\pi^0\eta)$ distribution equals the $\pi^0(\pi^+\eta)$ one. Absolute normalizations correspond to $A_1=1$ in Eq.~(\ref{ampls}). For two pions we give only the high mass $m_{\rm{high}}$ distribution.}
\label{figure3}
\end{figure}

We report in Fig.~\ref{figure3}, the theoretical curves for the phase space corrected probability, $|F_{K\bar K}|^2$ Eq.~(\ref{ratepiKK}), for $D_s \to \pi^+ K^+K^-$, $D_s \to \pi^+ K^0\bar K^0$ and  $D_s \to \pi^0 K^+\bar K^0$ as functions of $\sqrt{s}=m_{K\bar K}$. 
Curves are given for the central values of the parameters and for:
\be
Q=\frac{A_0}{A_1}=1.5
\label{Qmin}
\ee
which  minimizes the ratio of the rates (\ref{Kneutr}) to (\ref{Kcharg}), each integrated  between threshold and 1.15 GeV. The suppression of the neutral channel is evident. 

The value in (\ref{Qmin}) follows in fact from a very simple argument. Close to threshold we neglect the $s-M^2$ term as well as the kaon rates in the denominator of (\ref{sdepBW}), so that:
\be
i{\cal A}_{I=0}\simeq \left (\frac{g_0}{M_f\Gamma_0}\right)A_0;~~~~i{\cal A}_{I=1}\simeq\left(\frac{g_1}{M_a\Gamma_1}\right)A_1
\ee

Using the numerical values, we see that the factors in parentheses are in the ratio of about $1.3$ for $(I=1)/(I=0)$, so that the value (\ref{Qmin}) leads to nearly cancel the $K^0\bar K^0$ amplitude at threshold, see Eq.~(\ref{ispinampl})

With $Q$ fixed, we can predict the S-wave rates of $D_s\to 3\pi$ and  $D_s\to \eta2\pi$,  (\ref{Dspi0}), normalized to $\pi^+ K^+K^-$. 

The $3\pi$ and $\eta 2\pi$ decays are described each by two diagrams, differing by the exchange of $\pi^+\pi^+$ and $\pi^+\pi^0$, respectively.
We obtain the mass distribution by one-dimensional integration of the Dalitz-plot density, Eq.~(\ref{Dpdensity}). 
We give in Fig.~\ref{figure3}, right panel, the distribution for $D_s \to \pi^+(\pi^+\pi^-)$ and $D_s \to \pi^0(\pi^+\eta)$ as functions of the invariant mass of the particles indicated in parenthesis. The $D_s \to \pi^+(\pi^0\eta)$ distribution is equal to the $\pi^0(\pi^+\eta)$ one. 

In the limit of exact isospin (i.e. neglecting $f_0$-$a_0$ mixing) the $3\pi$ channel selects the $I=0$ scalars, $f_0(980)$ and $f_0(600)$. However the presence on one $s\bar s$ pair in the final state, Fig.~\ref{3qpairs}, implies that the $f_0(600)$ should be suppressed as assumed in Fig.~\ref{figure3}.

In conclusion, we report in Table 3 the ratios of the branching ratios of the $D_s$ modes considered thus far to the $\pi^+ K^+ K^-$ mode:
\be
R(X)=\frac{\rm{Br}(D_s\to X)}{\rm{Br}(D_s\to \pi^+ K^+K^-)}
\label{eq:tq}
\ee

\begin{table}[htb]
\caption{{\footnotesize The numerical values of the ratios $R(X)$, Eq.~(\ref{eq:tq}), for different values of $Q$}}
\begin{center}
\label{tab:3}
\begin{tabular}{@{}|c|c|c|c|c|}
\hline
$Q$ & $R(\pi^+ K^0\bar{K}^0)$ &  $ R(\pi^0 K^+\bar{K}^0) $ & $R(\pi^+\pi^-\pi^+)$ & $R(\eta\pi^0\pi^+)$  \\
\hline
 $1$ & $0.23 $ & $0.33$ &$ 1.12 $ & $8.9$\\
 \hline
 $1.5$ & $0.027 $ & $0.23$ &$ 1.74 $ & $6.15$\\
 \hline
 $2$ & $0.046 $ & $0.17$ &$ 2.3 $ & $4.51$\\
\hline
\end{tabular}\\[2pt]
\end{center}
\end{table}

\section{$D^0\to \bar K^0(K^+K^-)_{S-wave}$ and $D^0\to K^-(K^+\bar K^0)_{S-wave} $}

We consider these decays as arising from the quark process similar to those in Fig.~\ref{3qpairs}, with an $s\bar s$ pair taken from the sea:
\be
c+\bar u_{\rm spect}~\to~s+(u\bar d)+(s\bar s)+\bar u_{\rm spect}
\ee
The final state mesons are formed from quark states by different diagrams with respect to those considered for $D_s$ decays. This indicates that the amplitudes of $D_s$ and $D^0$ are not related by symmetry, even in the exact flavor-SU$_3$ limit. The weak hamiltonian~(\ref{heffect})  behaves under SU$_3$ according to: 
\be 
H_W=({\bar s}c)({\bar u}d)={\bf {6\oplus{\overline{15}} }}
\ee
and there are altogether seven independent couplings for the S-wave trasition $D\to $ Pseudoscalar +Scalar, with both S and P octet (eleven couplings, if we allow for singlets). The independent ways to couple the three quark pairs in Fig.~\ref{3qpairs} to the final $q\bar q$ octet meson are less (seven in all, including singlets), nonetheless there is no simple relations between the $D_s$ and $D^0$ couplings even in the quark diagram approximation.

\paragraph{The $K^-(K^+\bar K^0) $ channel.} The $K^-$ is formed from the spectator $\bar u$ plus a strange quark taken either from the weak decay or from the sea. The amplitude is not related by symmetry to the $D_s$ amplitudes previously found.  The system recoiling against the $K^-$ is dominated by the $a_0^+$. The dependence from $\sqrt{s}=m_{K\bar K}$ is well reproduced~\cite{BaBar1} with the parameters given in Table~\ref{tab:uno}.
\paragraph{The $K^0(K^+K^-)$ channel.} The $\bar K^0$ is formed from the current $\bar d$ plus a strange quark taken either from the weak decay or from the sea. In both cases, a ($su\bar s\bar u$) system is left, which should decay in $K^+K^-$ but not $K^0\bar K^0$. We encounter here the same situation that we analyzed in Sect.~\ref{analysis}. We are unable to relate by symmetry the $I=0$ and $I=1$ amplitudes of this decay, $A^\prime_{0,1}$,  to those of $D_s$ decays, eqs.~(\ref{extamp}). However, in the pole approximation, the conspiracy between the $f_0$ and $a_0$ terms must be the same as in~(\ref{Qmin}), to cancel the  $K^0\bar K^0$ rate.

\paragraph{The $K^0 K^0\bar K^0$ channel.} A clear prediction of the present scheme, is that this channel should be suppressed, if so is the $D_s\to \pi^+(K^0\bar K^0)$.

To compare with BaBar data~\cite{BaBar1}, we consider  the phase space corrected distributions, $|F_{K\bar K}|^2$ defined as in eq~(\ref{ratepiKK}), for the reactions (\ref{D03K}). From the previous discussion, we derive the forms:
\bea
&&A\left[D^0\to K^-(K^+\bar K^0)\right]={\rm C}^\prime~g_1~BW_a(s)\notag \\
&&A\left[D^0\to \bar K^0(K^+ K^-)\right]={\rm C}^{\prime\prime}\left[g_1~BW_a(s)+Qg_0~BW_f(s)\right]
\label{D3K}
\eea
with C$^\prime$ and C$^{\prime\prime}$ two independent constants and $Q$ given by (\ref{Qmin}).

\begin{figure}
\begin{center}
\includegraphics[width=0.45\textwidth]{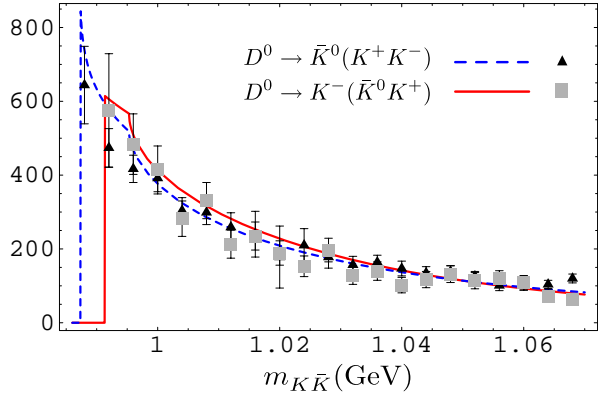}
\end{center}
\caption{\footnotesize Theory (for $Q=1.5$) and data (boxes and triangles~\cite{BaBar1}) as functions of $m_{K^+K^-}$, blue dashed, and $m_{K^+\bar K^0}$, continuous red (curves normalized to the data).
}
\label{fig:Swavedist}
\end{figure}
We report in Fig.~\ref{fig:Swavedist} theory and data as functions of $m_{K^+K^-}$, blue dashed, and $m_{K^+\bar K^0}$, continuous red (curves normalized to the data).

The agreement  is remarkable. In particular, it is to be noted the near coincidence of the $K^+\bar K^0$ and $K^+K^-$ distributions, which are dominated by $a_0^+$ and by a superposition of $f_0$ and $a_0^0$, respectively. This, of course, is due to the validity of the argument of Flatt\`e {et al.}~\cite{flatte}, similarly to the $D_s$ case.

\section*{Acknowledgements}
It is a pleasure to acknowledge interesting discussions with our colleagues of BaBar-Roma, in particular with G. Cavoto, F. Ferrarotto and M. Gaspero.



\begin{thebibliography}{99}
\bibitem{E791} 
  E.~M.~Aitala {\it et al.}  [E791 Collaboration],
  Phys.\ Rev.\ Lett.\  {\bf 86}, 765 (2001);
  {\it ibid.}, 770 (2001).
  
\bibitem{BaBar1}
  B.~Aubert {\it et al.}  [BABAR Collaboration],
  Phys.\ Rev.\  D {\bf 72}, 052008 (2005).
 
\bibitem{BESII} 
  M.~Ablikim {\it et al.}  [BES Collaboration],
  Phys.\ Lett.\  B {\bf 607}, 243 (2005).
 
 
 \bibitem{4qmesons} 
  L.~Maiani, F.~Piccinini, A.~D.~Polosa and V.~Riquer,
  Phys.\ Rev.\ Lett.\  {\bf 93}, 212002 (2004); {\it see also}
  arXiv:hep-ph/0604018;
   I.~Bigi {\it et al.} 
  Phys.\ Rev.\  D {\bf 72}, 114016 (2005).

  
  {\it Earlier references on multiquark mesons:} 
  R.~L.~Jaffe,
  Phys.\ Rev.\  D {\bf 15}, 281 (1977);
  H.~J.~Lipkin,
  Phys.\ Lett.\  B {\bf 70}, 113 (1977);
  G.~C.~Rossi and G.~Veneziano,
  Nucl.\ Phys.\  B {\bf 123}, 507 (1977);
   L.~Montanet, G.~C.~Rossi and G.~Veneziano,
  Phys.\ Rept.\  {\bf 63} (1980) 149;
  J.~D.~Weinstein and N.~Isgur,
  Phys.\ Rev.\ Lett.\  {\bf 48}, 659 (1982).

{\it The diquark-anti-diquark hypothesis was suggested in:} 
  R.~L.~Jaffe and F.~Wilczek,
  Phys.\ Rev.\ Lett.\  {\bf 91}, 232003 (2003).

   {\it Reviews on scalar mesons and exotic states:}
  F.~E.~Close and N.~A.~Tornqvist,
  ``Scalar mesons above and below 1-GeV,''
  J.\ Phys.\ G {\bf 28}, R249 (2002); 
  F.~E.~Close,
  ``Hadron spectroscopy (theory): Diquarks, tetraquarks, pentaquarks and no
  quarks,''
  Int.\ J.\ Mod.\ Phys.\  A {\bf 20}, 5156 (2005);
  R.~L.~Jaffe,
  ``Exotica,''
  Phys.\ Rept.\  {\bf 409}, 1 (2005)
  [Nucl.\ Phys.\ Proc.\ Suppl.\  {\bf 142}, 343 (2005)].

\bibitem{flatte} S. M. Flatt\`e {\it et al}., Phys. Lett. {\bf B38},  232 (1972).

\bibitem{sigma}  
  M.~Ablikim {\it et al.}  [BES Collaboration],
  Phys.\ Lett.\  B {\bf 598} (2004) 149.

{\it Recent theoretical support for the $\sigma$ resonance is found in:} I.~Caprini, G.~Colangelo and H.~Leutwyler,
  Phys.\ Rev.\ Lett.\  {\bf 96} (2006) 13200.

  {\it An account on the $q\bar{q}$ interpretation of scalar mesons can be found in:} 
  N.~A.~Tornqvist,
  ``Understanding the scalar meson $q \bar{q}$ nonet,''
  Z.\ Phys.\  C {\bf 68}, 647 (1995); see also
  A.~Deandrea, {\it et al.} 
  Phys.\ Lett.\  B {\bf 502}, 79 (2001);
  R.~Gatto {\it et al.} 
  Phys.\ Lett.\  B {\bf 494}, 168 (2000);
  A.~Deandrea and A.~D.~Polosa,
  Phys.\ Rev.\ Lett.\  {\bf 86}, 216 (2001).
  
 {\it On scalar mesons see also:}
  J.~R.~Pelaez,
  Phys.\ Rev.\ Lett.\  {\bf 92}, 102001 (2004);
  D.~Black, A.~H.~Fariborz, F.~Sannino and J.~Schechter,
  Phys.\ Rev.\  D {\bf 59}, 074026 (1999);
   M.~G.~Alford and R.~L.~Jaffe,
  Nucl.\ Phys.\  B {\bf 578}, 367 (2000).


\bibitem{kappa}  
        E.~M.~Aitala {\it et al.}  [E791 Collaboration],
        Phys.\ Rev.\ Lett.\  {\bf 89} (2002) 121801.
       


\bibitem{massmix} 
 L.~Maiani, F.~Piccinini, A.~D.~Polosa and V.~Riquer,
  Phys.\ Rev.\  D {\bf 70}, 054009 (2004); G.~C.~Rossi and G.~Veneziano,
  Phys.\ Lett.\  B {\bf 597}, 338 (2004).
  
 \bibitem{closemix} Small but non-negligible mixing was suggested, from central production data, in: F.~E.~Close and A.~Kirk,
  Phys.\ Lett.\  B {\bf 489} (2000) 24.

\bibitem{CrystBarr} 
  A.~Abele {\it et al.} [Crystal Barrel Collaboration]
  Phys.\ Rev.\  D {\bf 57}, 3860 (1998).
 
 
\bibitem{PDG}
  W.~M.~Yao {\it et al.}  [Particle Data Group],
  J.\ Phys.\ G {\bf 33} (2006) 1.
  

\end{thebibliography}
\end{document}